\documentclass{mn2e}
\usepackage{psfig}
\usepackage{longtable}

\def\ltsima{$\; \buildrel < \over \sim \;$}
\def\lsim{\lower.5ex\hbox{\ltsima}}
\def\gtsima{$\; \buildrel > \over \sim \;$}
\def\gsim{\lower.5ex\hbox{\gtsima}}

\begin{document}

\title[Bulk Comptonization spectra in blazars]
{Bulk Comptonization spectra in blazars}

\author[Celotti, Ghisellini and Fabian] {A. Celotti$^1$, G.
  Ghisellini$^2$ and A.C. Fabian$^3$\\ $^1$ SISSA/ISAS, Via Beirut
  2-4, I-34014, Trieste, Italy {\tt e-mail: celotti@sissa.it}\\ $^2$
  INAF, Osservatorio Astronomico di Brera, via E. Bianchi 46, I-23807
  Merate (LC), Italy {\tt e-mail: gabriele@merate.mi.astro.it}\\ $^3$
  Institute of Astronomy, Madingley Road, CB3 0HA Cambridge {\tt
    e-mail: acf@ast.cam.ac.uk}}

\maketitle

\begin{abstract}
  We study the time dependent spectra produced via the bulk Compton
  process by a cold, relativistic shell of plasma moving (and
  accelerating) along the jet of a blazar, scattering on external
  photons emitted by the accretion disc and reprocessed in the broad
  line region.  Bulk Comptonization of disc photons is shown to yield
  a spectral component contributing in the far UV band, and would then
  be currently unobservable.  On the contrary, the bulk Comptonization
  of broad line photons may yield a significant feature in the soft
  X--ray band. Such a feature is time--dependent and transient, and
  dominates over the non thermal continuum only when: a) the
  dissipation occurs close to, but within, the broad line region; b)
  other competing processes, like the synchrotron self--Compton
  emission, yield a negligible flux in the X--ray band.  The presence
  of a bulk Compton component may account for the X--ray properties of
  high redshift blazars that show a flattening (and possibly a hump)
  in the soft X--rays, previously interpreted as due to intrinsic
  absorption.  We discuss why the conditions leading to a detectable
  bulk Compton feature might be met only occasionally in high redshift
  blazars, concluding that the absence of such a feature in the
  spectra of most blazars should not be taken as evidence against
  matter--dominated relativistic jets. The detection of such a
  component carries key information on the bulk Lorentz factor and
  kinetic energy associated to (cold) leptons.
\end{abstract}


\section{Introduction}

High redshift blazars are the most powerful observed radio--loud AGN,
and among the best candidates to be detected by the forthcoming GLAST
$\gamma$--ray satellite.  Recently, one previously unidentified EGRET
source has been associated with the blazar Q0906--693 at a redshift
5.47 (Romani et al. 2004). Its spectral energy distribution (SED)
resembles the predictions of the proposed `blazar sequence' (Fossati
et al. 1998; Ghisellini et al. 1998), namely the presence of two broad
peaks in the mm--far IR and the MeV--GeV bands, with the high energy
component strongly dominating the energy output.  Furthermore, there
are now half a dozen blazars at $z>4$, yet undetected in the MeV--GeV
band, but observed in the X--ray band with large area instruments
(i.e. XMM--{\it Newton}, see Yuan et al. 2006).  All of them exhibit a
flat (i.e. rising in $\nu F_\nu$) spectrum, indicating the energetic
dominance of a spectral component peaking at high frequencies.  The
inferred apparent luminosities reach $> 10^{49}$ erg s$^{-1}$, yet the
intrinsic ones (i.e. corrected for the effect of relativistic beaming)
are likely to be only a minor fraction of the jet power which
energizes the radio lobes.

Despite the theoretical advances following the discovery of the
radiatively dominant $\gamma$--ray emission in blazars, and more than
thirty years after the discovery of superluminal motion in 3C~273
(Whitney et al. 1971; Cohen et al. 1971) we are still unable to pin
down the acceleration and collimation mechanism(s) and establish
whether the energy carried in jets is mainly in the form of Poynting
flux or kinetic energy of matter and how this might depend on the
distance from the powering source (see the arguments presented by
Blandford 2002 and Lyutikov 2003 in favour of
electromagnetically--dominated jets, and Sikora et al. 2005 for a
discussion of matter vs magnetic jets).

One of the proposed diagnostics on the issue of jet composition (e.g.
Sikora \& Madejski 2000) is based on the spectral signature expected
from a matter--dominated jet, in the form of an excess of emission at
around 1 keV, due to the bulk Comptonization by cold leptons of the
radiation fields produced by the accretion disc and/or reprocessed in
the broad line region (BLR).  This process, first suggested by
Begelman \& Sikora (1987), was then considered by Sikora, Begelman \&
Rees (1994) and Sikora et al. (1997).  Moderski et al. (2004) proposed
that such a process can generate a soft X--ray precursor in the light
curve of blazars.

No evidence for bulk Comptonization features have been detected so
far, leading to upper limits on the jet matter content (see above
references). However, there is increasing recent evidence that the
X--ray spectrum of some high redshift blazars show a flattening
towards low X--ray energies and possibly a hump.  The favoured
interpretation so far has been the presence of intrinsic absorption by
warm material, but alternative explanations are viable (see Fabian et
al. 2001b, Worsley et al. 2004). In particular it is possible that for
the first time we are observing the signature of the bulk
Comptonization process.

This motivates the present work which concentrates on the
(time--dependent) detailed shape of the bulk Compton spectrum to
assess this hypothesis against available and future data, under
reasonable assumptions about the presence of ambient seed photons and
the bulk acceleration of the plasma in the jet.  Future X--ray
missions could provide variability as well as polarimetric information
in the X--ray band, the latter being a further valuable tool to single
out emission via bulk Comptonization among competing radiation
processes (Begelman \& Sikora 1987), such as synchrotron self--Compton
(SSC) (Celotti \& Matt 1994), and external Compton (Poutanen 1994).

\section{Model Assumptions}

We consider a shell comprising $N$ cold (i.e. non relativistic)
leptons moving along a jet, accelerating up to a saturation speed.
During propagation, the leptons scatter ambient photons, originating
both in a (standard) radiatively efficient accretion disc and
reprocessed in the gas clouds forming the BLR, as expected in powerful
quasars.  In the following we detail the assumptions made in this
scenario.

\subsection{Plasma dynamics}

The plasma dynamics, and in particular the dependence of the jet
velocity on distance $z$ from the power source, is not known. Modeling
of magnetically--accelerated flows leads to relatively slow
acceleration (e.g. Begelman \& Li 1994), before the flows reaches an
asymptotic speed. Here, we parametrize the acceleration in terms of
the bulk Lorentz factor $\Gamma$ increasing with distance (i.e. the
height above the disc) as:
\begin{equation}
\Gamma(z) \, = \, \Gamma_0 \left( {z\over z_0}\right)^a,
\label{g}
\end{equation}
up to a maximum value $\Gamma_{\rm max}$, beyond which
$\Gamma=\Gamma_{\rm max}$. The initial bulk Lorentz factor, formally a
free parameter, will be considered to be 1, unless otherwise
specified.  The initial height $z_0$ (corresponding to the typical
distance where acceleration sets in) will be taken of the same order
as the inner accretion radius $r_0$.  To simplify the notation, we
hereafter set $\Gamma=\Gamma(z)$ and
$\beta=\beta(z)=(\Gamma^2-1)^{1/2}/\Gamma$.

\subsection{Scattering of radiation from the accretion disc}

To model the radiation produced by the accretion disc, we simply
consider the surface (multi-colour blackbody) temperature profile for
a Shakura \& Sunyaev (1973) disc:
\begin{equation}
T_{\rm D}(r) \, \propto \, M_{\rm BH}^{-1/2} \dot M^{1/4} r^{-3/4}
\left[ 1- \left({6r_g\over r}\right)^{1/2} \right]^{1/4},
\label{tr}
\end{equation}
where $r$ is the radial disc coordinate, $r_g=GM_{\rm BH}/c^2$ the
gravitational radius, $M_{\rm BH}$ the central black hole mass and
$\dot M$ is the mass accretion rate.

At a given height $z$, the bolometric radiation energy density
produced by a `ring' of the accretion disc, as seen in the lab frame,
corresponds to
\begin{equation}
{dU(z)\over dr} \, = \, {2\pi r B(T_{\rm D}) \cos\theta\over c
(z^2+r^2)}.
\label{dupdr}
\end{equation}
$B(T_{\rm D})=\sigma_{\rm SB} T_{\rm D}^4 /\pi$ represents the
frequency--integrated black body intensity, $\sigma_{\rm SB}$ is the
Stefan-Boltzmann constant, and $\cos\theta= [1+(r/z)^2]^{-1/2}$, 
$\theta$ being the angle
between the position $z$ along the jet and the emitting ring of the
disc.
In the comoving (primed) frame of the shell this radiation energy
density is given by:
\begin{equation}
{dU^\prime(z)\over dr} \, = \Gamma^2(1-\beta\cos\theta)^2\,
{dU(z)\over dr}.
\label{dudr}
\end{equation}
We assume that the scattering process is isotropic, i.e. we neglect
the angle dependence of the differential Thomson cross section, by
setting $d\sigma/d\Omega =\sigma_{\rm T}/(4\pi)$.  Thus, for an
optically thin shell (see Sect. 2.5), each electron scatters a
fraction $\sigma_{\rm T}$ of this radiation energy density,
re--isotropizing it in the comoving frame. For $N$ free and cold
electrons the scattered radiation (at a given $z$) corresponds to a
bolometric luminosity (in the comoving frame)
\begin{equation}
{dL^\prime(z)\over dr} \, = \sigma_{\rm T} c N \, {dU^\prime(z)\over
dr}.
\label{dlpdr}
\end{equation}
An observer located at an angle $\theta_{\rm V}$ with respect to the
jet axis will receive a power
\begin{equation}
{dL_{\rm obs}(z)\over dr} \, = {1\over [\Gamma (1-\beta\cos\theta_{\rm V})]^4}
{dL^\prime(z)\over dr} \, \equiv\, \delta^4(z){dL^\prime(z)\over dr},
\label{dldr}
\end{equation}
where $\delta=\Gamma^{-1} (1-\beta \cos\theta_{\rm V})^{-1}$ defines
the relativistic Doppler factor.
In terms of the spectrum, the observed radiation will still have a
blackbody frequency distribution, corresponding to a transformed
temperature:
\begin{equation}
T_{\rm D, obs}(r, z) \, =\, T_{\rm D}(r) \, { 1-\beta\cos\theta
\over 1-\beta\cos\theta_{\rm V}}.
\label{tobs}
\end{equation}

Given the total luminosity (eq.~6) and spectrum (eq.~7), the observed
spectrum can be normalized by setting
\begin{equation}
{dL_{\rm obs}(z, \nu)\over dr} \, =\, A \, {2h \over c^2}\, {\nu^3 \over
 \exp(h\nu/kT_{\rm D,obs}) -1},
\label{bb_obs}
\end{equation}
\begin{equation}
A \, =\, { 2\pi \sigma_{\rm T} N r \cos\theta \over z^2 +r^2} \,
{1\over [\Gamma (1-\beta\cos\theta)]^2 }.
\label{A}
\end{equation}
The observed power emitted at each $z$ can be simply obtained by
integrating over the disc radii $r$, between $r_0$ and an outer disc
radius (see Sec.~2.5).

In the following we will also consider the time integrated spectrum,
as measured by the observer over an (integration) time $t_{\rm obs}$,
given by:
\begin{equation}
dt_{\rm obs} = {dz \over \beta c}\, (1-\beta \cos\theta_{\rm V}),
\label{dt}
\end{equation}
leading to
\begin{equation}
t_{\rm obs} = \int_{z_0}^z {dz^\prime \over \beta c}\, 
(1-\beta \cos\theta_{\rm V}).
\label{dt}
\end{equation}

\subsection{Scattering of broad line photons}

Besides the photons directly originating in the disc, a significant
contribution to the soft photon field can come from disc photons
reprocessed by the gas permeating the BLR (Sikora, Begelman \& Rees
1994). We account for this contribution assuming that the photon
energy distribution follows a blackbody spectrum peaking at the
frequency of the Lyman-$\alpha$ hydrogen line,
$\nu_{L\alpha}=2.47\times 10^{15}$ Hz.  This matches the shape of this
external radiation component in a restricted energy range around the
Lyman--$\alpha$ line, as seen in the comoving frame.

In fact for BLR clouds distributed in two semi--spherical shells (one
for each side of the accretion disc), any monochromatic line is seen,
in the comoving frame, within a narrow cone of semi-aperture angle
$1/\Gamma$ along the jet velocity direction.  Accordingly, such
photons are blueshifted by a factor ranging from $\Gamma$ (photons
from the border of the cone) to $2\Gamma$ (photons head--on).  In this
(admittedly narrow) frequency range a monochromatic line transforms
into a spectrum $\propto {\nu_{\rm obs}}^2$.  Although the
Lyman--$\alpha$ line represents the most important contribution of the
BLR seed photons, the entire spectrum produced by the BLR is more
complex (see Tavecchio et al., in prep.).  Here, we stress that the
blackbody assumption represents well the more complex SED from the
BLR.

Thus we set an equivalent temperature of the BLR spectrum: 
\begin{equation}
T_{\rm BLR}\, \equiv\, { h\nu_{L\alpha} \over 2.8 k} \,\sim\, 4.23
\times 10^4\, {\rm K}.
\end{equation}
In the shell comoving frame this temperature is seen blueshifted by a
factor $\sim\Gamma$, while, after scattering, the transformation in the
observer frame introduces an additional blueshift, corresponding to a
factor $\delta$: therefore $T_{\rm BLR, obs} \sim \Gamma\delta T_{\rm BLR}$.

The bolometric luminosity resulting from the scattering of BLR
photons, at a given distance $z$, is given by
\begin{equation}
L_{\rm BLR, obs}(z)\, = \, \sigma_{\rm T} c N U_{\rm BLR} \Gamma^2
\delta^4,
\end{equation}
where $U_{\rm BLR}$ is the energy density of BLR photons, which is
considered constant within the typical distance of the BLR, defined by
a radius $R_{\rm BLR}$.  For a luminosity $L_{\rm BLR}$ of broad line
photons, the observed spectrum will have a blackbody shape,
corresponding to the temperature $T_{\rm BLR, obs}$:
\begin{eqnarray}
L_{\rm BLR, obs}(z, \nu)\, &=& \, \, { 2h \over c^2} \,\, {\nu^3 \over
\exp\left(h\nu/kT_{\rm BLR, obs}\right) -1} \, \times \nonumber \\ &~& {
\sigma_{\rm T} N L_{\rm BLR} \over 4\sigma_{\rm SB}\, R^2_{\rm BLR}
T_{\rm BLR}^4 \Gamma^2}
\end{eqnarray}
which corresponds to the integrated luminosity given by Eq.~13.

\subsection{Disc vs broad line photons}

Since the highest temperature is reached in the innermost parts of the
accretion disc, the angle between the energetic disc photons and the
moving shell is typically small.  It is relatively large only in the
vicinity of the disc, but there the shell has not yet reached high
bulk Lorentz factors.  As a consequence (see Eq.~7), the observed
typical frequency of the scattered disc photons will not increase much
(for $\theta\sim 0$ and $\theta_{\rm v}\sim \beta$ such photons are
even redshifted).

On the contrary, photons from the BLR are always seen head--on, and as
such are maximally blueshifted (i.e. by the factor $\Gamma\delta$).
For this reason (as shown below) it will be this bulk Compton
component which can provide a relevant contribution in the X--ray
spectrum of powerful blazars. 

\subsection{Model parameters}

In the following we quantitatively specify the choice of model
parameters adopted in the spectral calculations presented in 
Section~3. 

\vskip 0.3 true cm
\noindent
{\bf Disc luminosity ---} For powerful blazars we assume a typical
black hole mass $M_{\rm BH}=10^9 M_9 M_{\odot}$ and a disc accretion
rate close to the Eddington one. Therefore
\begin{equation}
L_{\rm disc} \, \sim \, L_{\rm Edd}\, \sim 10^{47} M_9 
\,\, {\rm erg\,\, s^{-1}},
\end{equation}
and the typical sizes of the inner accretion disc and jet are assumed
$z_0=r_0=6 r_{\rm g}$.  The outer accretion disc radius is set to
$10^4$ $r_{\rm g}$.

\vskip 0.3 true cm
\noindent
{\bf Broad line radiation ---} We model the broad line region as a
thin spherical shell at distance $R_{\rm BLR}$. The disc radiation
reprocessed by clouds in the BLR is assumed to have a bolometric
luminosity $L_{\rm BLR}=0.1 L_{\rm disc}$.

It should be noted that the BLR may not be the only contributor to
seed photons, besides those directly from the accretion disc.  A
molecular torus (emitting infrared photons; e.g. Blazejowski et al. 2004), 
and intra--cloud plasma (scattering and isotropizing both disc
and also beamed synchrotron jet radiation; e.g. Ghisellini \& Madau,
1996) may also provide soft photons. However we neglect such
components here, as their contribution is very uncertain.

\vskip 0.3 true cm
\noindent {\bf Number of electrons---} We consider a shell of plasma,
of some initial size equal to the inner size of the accretion disc,
$r_{\rm 0}$.  As first order estimate on how many particles can be
present in such a shell, we consider the number of {\it relativistic}
electrons needed to originate the radiation dominating the SED (i.e.
produced in the jet regions where most of the dissipation occurs) and,
assuming that the total number of particles is conserved, we treat it
as a {\it lower limit} to the total number of electrons present in the
shell.

This limit is clearly model dependent. We will adopt the most common
interpretation that the high energy radiation in the spectrum of
powerful blazars is produced via inverse Compton emission of a
relativistic distribution of leptons on the external photon fields
discussed in Sec. 2.2 and 2.3. In this scenario the inferred number
density of relativistic particles relies on estimates of the magnetic
field and seed photons energy density required to give raise to the
observed luminosity.  On the other hand, these quantities are quite
tightly constrained by fitting the whole SED -- both the synchrotron
and inverse Compton components. Furthermore for powerful blazars, the
electron distribution is basically in the complete cooling regime
(namely, electrons of all energies cool within a light crossing time),
which constrains the low energy particle end.

For a reference value of $N=10^{54}N_{54}$ particles (see Sect. 3),
which initially occupy a volume of typical size $r_0 = f 2 r_{\rm g}$
($2 r_{\rm g}$ being the Schwarzschild radius), the corresponding
initial optical depth is $\tau_0 \equiv \sigma_{\rm T} r_0 (N/r_0^3)
\sim 7.4 N_{54}/(f M_9)^2$.
At the base of the jet, the shell can be optically thick for
scattering (depending on $f$): in the calculations we rescale the bulk
Compton emission by the factor max$(1,\tau)$, which accounts for the
reduction in the scattered flux when $\tau>1$ (self--shadowing).
However, for a conical jet the region rapidly becomes transparent as
$\tau \propto z^{-2}$, independently of the shell width. After
transparency is reached all electrons contribute to the emission and
the process becomes largely independent of the specific geometry
(shell or spherical), and the expressions given in Sect. 2.4 are
appropriate.

\vskip 0.3 true cm
\noindent {\bf Plasma dynamics ---} As mentioned above we parametrize
the uncertain dynamics as $\Gamma(z)=(z/z_0)^a$. In the following
examples we adopt $a=1/2$, which, for $\Gamma_{\bf max}\sim 10$,
corresponds to reach the asymptotic bulk Lorentz factor at distances
$z=z_0\Gamma_{\rm max}^2 \sim$ a few hundreds $r_{\rm g}$.  In the
context of the internal shock scenario (Rees 1978; Sikora, Begelman \&
Rees 1994; Ghisellini 1999; Spada et al. 2001) this is approximately
the distance where shell--shell collisions take place, i.e. where most
of the dissipation occurs.  For $a=1$ (i.e. more rapid acceleration)
the results are qualitatively the same, but the observed time
evolution is quicker.

\begin{figure}
\vskip -0.8 true cm
\psfig{figure=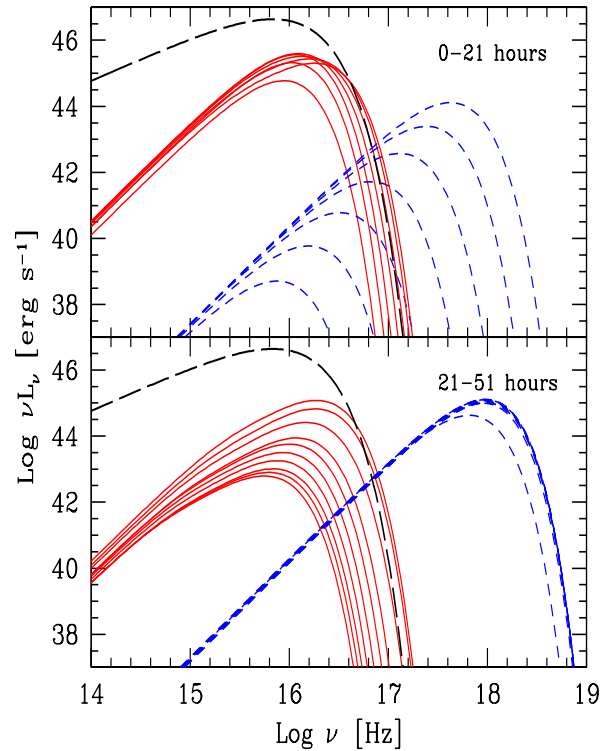,width=9cm,height=12cm}
\vskip -0.8 true cm
\caption{Temporal evolution of the bulk Compton spectrum. Long dashed
  line: accretion disc spectrum.  Solid lines: bulk Compton spectra
  from scattered disc photons.  Dashed lines: bulk Compton spectra
  from scattered broad line photons.  Each spectrum corresponds to the
  average luminosity, assuming 3 hours of integration time, produced
  by the shell as it moves from $z_0$ to the BLR, assumed to be at
  $R_{\rm BLR}= 10^{18}$ cm.  The top panel shows the evolution for
  the first 21 hr (observer time) during which the contribution from
  bulk Comptonization of disc photons increases; the bottom panel
  corresponds to the time interval between 21 and 51 hr, during which
  the same component decreases in flux.  Notice that the luminosity
  and spectrum from the bulk Comptonization of BLR photons remain
  constant after $\sim$ 21 hr (bottom panel), corresponding to the
  fact that $\Gamma$ has reached its maximum value.}
\end{figure}
\begin{figure}
\vskip -0.2 true cm
\psfig{figure=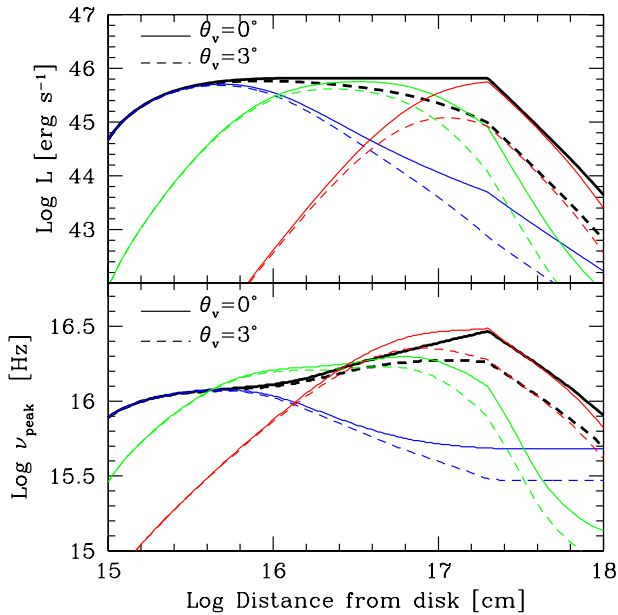,width=9cm,height=9cm}
\vskip -0.2 true cm
\caption{The bolometric bulk Compton luminosity from the scattering
  of disc photons (top panel) and the frequency peak of this component
  (bottom panel) as a function of the height of the shell above the
  disc, $z$.  Broad line photons are neglected here.  Thick (black)
  lines show the contribution from soft photons produced in the entire
  disc, while thin lines represent, respectively, the contribution
  from 6--60 $r_{\rm g}$ (blue); 60--600 $r_{\rm g}$ (green) and
  600--6000 $r_{\rm g}$ (red). Note that the three regions of the disc
  similarly contribute to the total luminosity.  Solid and dashed
  lines correspond to viewing angles $\theta_{\rm v}=0^\circ$ and
  $\theta_{\rm v}=3^\circ$, respectively. We assumed $L_{\rm
    disc}=10^{47}$ erg s$^{-1}$, $M_{\rm BH}=10^9 M_\odot$.}
\end{figure}

\section{Results}

Fig. 1 shows a sequence of bulk Compton spectra resulting from
scattering disc and line photons.  The sequence refers to spectra at
different times (in the observer frame) from the beginning of the
plasma acceleration at $z_0=6r_{\rm g}$.  Each spectrum corresponds to
the average luminosity received in 3 hr of integration time, $t_{\rm
  obs}$, again as measured in the rest frame of the source.  The
acceleration ($\Gamma\propto z^{1/2}$) terminates when $\Gamma_{\rm
  max}=15$ is reached.  For comparison the BLR size is $R_{\rm
  BLR}=10^{18}$ cm $\sim 10^4 r_{\rm g}$.

The total number of cold particles is set at $N\sim 10^{53}$, and the
observer is located at a viewing angle $\theta_{\rm v}=3^\circ$.

The figure shows separately the contribution from scattering of disc
(continuous lines) and BLR (dashed lines) photons.  For clarity the
top panel shows the evolution of the spectrum in the rising phase of
the disc--scattered component (roughly until the Lorentz factor
reaches $\Gamma_{\rm max}$), while the bottom panel shows its decline.

Such evolution corresponds to the following phases:

\begin{itemize}
\item At early times, the bulk Lorentz factor is small, with a
  correspondingly low scattered luminosity, from both disc and line
  photons. The former component is more prominent as the radiation
  energy density of the disc photons is larger at small $z$.  The
  different slope of the disc--scattered radiation with respect to the
  multi--color black body is due to the dilution of photons produced
  at larger disc radii, weakly contributing to the total energy
  density.

\item As the jet accelerates (before reaching $\Gamma_{\rm max}$), the
  BLR--scattered component increases more rapidly than the
  disc--scattered one.  This is because the radiation energy density
  of the BLR photons is constant for a stationary observer inside it,
  but increases like $\Gamma^2$ in the shell comoving frame.  The
  radiation energy density produced by the inner disc, instead,
  decreases for a stationary observer, and even more so in the shell
  frame (since the photon directions form small angles with respect to
  the shell velocity).  Only photons produced at large disc radii are
  blueshifted, and make up the dominant part of the disc radiation
  energy density.

\item When the shell reaches its maximum Lorentz factor (at $t_{\rm
    obs} \sim 21$ hr), it is already relatively far from the disc. In
  the comoving frame a larger and larger portion of the disc is seen
  at small angles (i.e. in the comoving frame photons are seen
  redshifted).
 
  Correspondingly, the disc--scattered radiation intensity decreases
  (bottom panel of Fig.~1), while the BLR line--scattered component
  remains constant, and dominates.

\item As mentioned in Sec. 2.4, the typical energies of the disc and
  BLR line scattered components are different.  The disc--scattered
  hump peaks in the far UV, in the spectral region currently most
  difficult to observe.  This is because energetic disc photons reach
  the shell at small angles, except at very early times when however
  the Lorentz factor is still low.  Therefore they are not
  significantly boosted.  Line photons, instead, always intercept the
  shell head on (in the comoving frame), and thus are maximally
  boosted.  The peak frequency of this component depends only on the
  value of $\Gamma$ that the shell has at a given $z$ (and of course
  on the viewing angle).

\end{itemize}

To investigate the last item better, which is crucial in assessing the
detectability of bulk Compton emission, we show in Fig.~2 the
bolometric luminosity and the peak frequency of the disc--scattered
spectrum as a function of the shell--disc distance, $z$.  Solid and
dashed lines correspond to $\theta_{\rm v}=0^\circ$ and $\theta_{\rm
  v}=3^\circ$, respectively.  The contribution of seed photons from
the inner (6--60 $r_{\rm g}$), central (60--600 $r_{\rm g}$) and outer
(600--6000 $r_{\rm g}$) parts of the disc are shown separately: the
inner part is relevant only when the shell is close to the disc, while
the other part mostly contributes when the shell moves beyond a few
($\sim 10$) $r_{\rm g}$.  The curves change behavior when $\Gamma$
reaches $\Gamma_{\rm max}$ (or $z > R_{\rm BLR}$), which in this
particular example occurs at $z\sim 2\times 10^{17}$ cm.

\begin{figure}
\vskip -0.5 true cm
\centerline{\psfig{figure=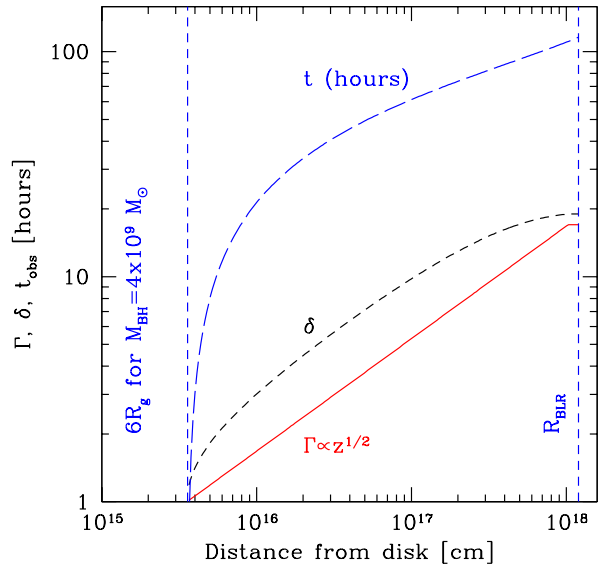,width=9cm,height=9cm}}
\vskip -0.6 true cm
\caption{The bulk Lorentz factor $\Gamma$, Doppler factor $\delta$,
  and observed (rest frame) time $t_{\rm obs}$ (the latter expressed
  in hours), as a function of the location of the shell above the
  disc. The input parameters are the same used in the previous case,
  except for $ L_{\rm disc}=4\times 10^{47}$ erg s$^{-1}$ (for a black
  hole of $M_{\rm BH}= 4\times 10^9 M_{\odot}$); $R_{\rm
  BLR}=1.2\times 10^{18}$ cm; a total number of (cold) particles
  $N=8.6\times 10^{54}$ and a viewing angle $\theta_{\rm
  v}=3^\circ$. The vertical dashed lines corresponds to the start of
  our calculations (i.e. $6 r_{\rm g}$) and the assumed location of
  the BLR.}
\end{figure}
\begin{figure}
\vskip -0.8 true cm
\centerline{\psfig{figure=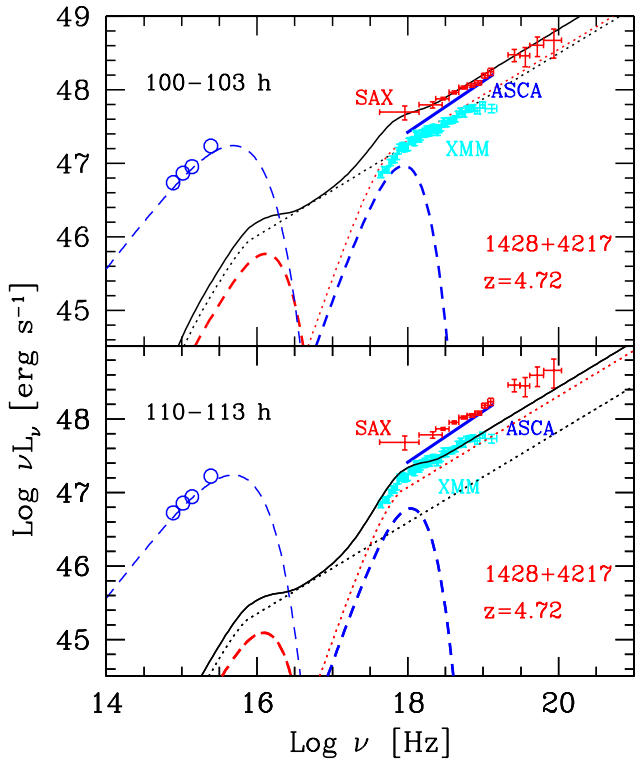,width=11.5cm,height=12cm}}
\vskip -0.6 true cm
\caption{Illustration of how the bulk Compton process can explain some
  details of the X--ray spectrum of a powerful blazars, GB B1428+4217.
  The X--ray data are from {\it Beppo-SAX} (red), XMM-{\it Newton}
  (cyan) and ASCA (blue).  The power law is assumed to start from the
  top of the two Comptonization humps, corresponding to assume that a
  number of leptons similar to those 'cold' are accelerated to
  relativistic energies and form a power law distribution extending
  down to $\gamma_{\rm min}=1$.  Top panel: spectrum predicted during
  3 hours of exposure time starting 100 hours after the beginning of
  the acceleration.  Bottom panel: the same, but after 110 hours. Due
  to the hardness of the power law, the contribution to the total
  power law of disc and broad line photons is almost equal even after
  100 hours, despite the fact that the bulk Compton hump from disc
  photons is less powerful than the one from the broad line
  photons. All frequencies are in the rest frame of the source.}
\end{figure}

\subsection{The case of GB B1428+4217}

To illustrate the possible relevance of the bulk Compton process for
the interpretation of the X--ray spectra of blazars, which motivates
this work, we consider the case of GB B1428+4217, a powerful blazar at
redshift $z=4.72$.  This is one of the best studied high redshift
($z>4$) blazars, the spectrum of which has been observed in X--rays by
ROSAT (Boller et al. 2000), {\it Beppo}SAX (Fabian et al. 2001a) and
XMM--{\it Newton} (Worsley et al. 2004; Worsley et al. 2006).  These
observations showed that the X--ray spectra flatten towards low
energies with respect to a higher energy power law.  This behaviour
has been interpreted as likely due to (warm) absorbing material (with
columns exceeding $10^{22}$ cm$^{-2}$).  Alternative interpretations
are however possible (Fabian et al. 2001b): a flattening of the low
energy part of an inverse Compton component could be due to a low
energy cut--off in the electron population and/or a sharply peaked
soft (seed) photon distribution.  The sharpness of the X--ray break
(occurring over a range of only a few keV in the rest--frame) is
however difficult to explain within these models.

Here we consider the possibility that the flattening and (possibly) a
hump in the soft X--ray spectrum (see Worsley et al. 2004, Yuan et al.
2005) reveal the presence of emission via bulk Comptonization.  For
this specific example we consider the parameters which allow to
reproduce the high energy spectral component on the basis of the
external inverse Compton process (e.g. in Ghisellini, Celotti \&
Costamante 2002). The SED can be satisfactorily fitted as emission of
a shell with cross section radius $\sim 2\times 10^{16}$ cm and width
(as measured in the comoving frame) $\Delta R^\prime=1.2\times
10^{15}$ cm, moving with a bulk Lorentz factor $\Gamma=17$. The
inferred magnetic field is $B=11$ G and the (comoving) number density
of relativistic particles $\sim 5.9\times 10^5$ cm$^{-3}$,
corresponding to a total number of relativistic particles $N_{\rm
  rel}\sim 8.8\times 10^{53}$.  The observer is at $\theta_{\rm
  v}=3^\circ$.

To account also for the soft X--ray flattening and hump, we assume
that again the plasma accelerates from $z_0=6r_{\rm g}$ according to
$\Gamma=(z/z_0)^{1/2}$, up to $\Gamma_{\rm max}=17$.  The dynamics of
the shell with respect to the observer time corresponds to the case
presented in Fig.~3.  The disc luminosity is $L_{\rm disc}=4\times
10^{47}$ erg s$^{-1}$, for a $M_{\rm BH}= 4\times 10^9 M_{\odot}$.
The BLR, located at $R_{\rm BLR}=1.2\times 10^{18}$ cm, reprocesses 10
per cent of the disc radiation.  The resulting spectra are shown in
Fig. 4, where we report different X--ray data sets, and optimize the
parameters to reproduce the {\it Beppo}SAX (top panel) and XMM-{\it
Newton} (bottom panel) states.

The overall SED clearly depends on the relative normalisation of the
relativistic and cold electron distributions. The power law components
in Fig.~4 (dotted lines) extend from the peaks of the two bulk Compton
components: this corresponds to a similar number of electrons in the
two particle distributions \footnote{For the relativistic particle
  distribution we assume $N(\gamma) \propto \gamma^{-(2\alpha+1)}$
  extending down to $\gamma_{\rm min}\sim 1$, where $\alpha$ is the
  spectral index, $L(\nu)\propto \nu^{-\alpha}$.}, namely a total
number of (cold) particles $N=4.0 (1.3)\times 10^{54}$ (for the top
and bottom panels, respectively).

The normalizations of the relativistic and cold particle distributions
have been assumed to be equal, and the two populations have been
considered independently. Clearly, a self-consistent treatment should
in principle account for the acceleration and cooling processes to
determine the relative normalization (in number and energy) as a
function of time. The assumption made here corresponds to about half
of the particle present at the dissipation site to be accelerated.
Alternatively this could correspond to a situation where all leptons
are cold for half of the exposure time, and relativistic for the other
half.

The spectra shown in Fig.~4 refer to the average luminosity detected
in three hr of integration time (rest frame), starting at $t_1$=100 hr
(top panel) and $t_2=$110 hr (bottom panel) with respect to the
beginning of the acceleration ($z=z_0$).  These times correspond to
locations of the shell quite close to $R_{\rm BLR}$ (see Fig.~3).

After $t_1$, the luminosities ($\nu L_\nu$) of the bulk Compton
spectra of disc and line-scattered photons are similar, while at $t_2$
the contribution from the line photons dominates. In fact, by $t_2$
the shell has already reached its maximum Lorentz factor, quite far
from the inner parts of the disc.

As already shown (see Fig.~1), the bulk Compton spectrum of the disc
photons, at all times, peaks in the far UV, and is thus undetectable.
However, its luminosity is highly time--dependent (see the top panel
of Fig.~1), exceeding the bulk Compton luminosity off line photons at
times $t\lsim t_1$ (rest frame) for the specific example shown here.

As a consequence should dissipation -- namely acceleration of
electrons to relativistic energies -- occur at lower height above the
disc, we should have a non--thermal spectrum dominated by the
scattering of the disc, not line, photons.  This would imply that the
X--ray excess due to bulk Compton off the line photons would become
undetectable, being exceeded by non--thermal radiation.  This can be
seen by comparing the spectra in the top and bottom panels of Fig.~4.
Early dissipation, however, faces a severe difficulty in accounting
for the SED, since the source becomes opaque to the $\gamma$--$\gamma
\to e^\pm$ process. The emission of the created pairs would contribute
at frequencies between the two broad peaks of the SED, resulting in an
overproduction of X--rays with respect to the observed level
(Ghisellini \& Madau 1996).

Note that -- as shown in Fig.~4 -- the power law continuum from
disc-scattered photons dominates over a limited frequency range even
if the corresponding bolometric energy density is smaller than that of
line-scattered photons. For non thermal scattered radiation with a
hard spectrum ($\alpha<1$) the typical frequencies of the seed photons
are important: disc radiation should not be discarded as main
contributor to the seed photon field, on the basis of its bolometric
energy density only (see Dermer \& Schlickeiser 1993).

In the observationally interesting X--ray range (i.e. 0.1--10 keV,
that translates into 0.5--50 keV rest frame for sources at $z\sim$ 4),
the spectrum can be approximated as the sum of a power law and a
blackbody. The latter component would correspond to a hump in the mid
X--ray band, accompanied by a flattening towards soft energies.

The bulk Compton feature is expected to be time--dependent, and
generally transient.  The necessary conditions for it to be detectable
are: i) dissipation must occur far from the accretion disc, but within
and near to the BLR; ii) the shell should move close to its maximum
speed; iii) other competing processes, such as SSC, should not
dominate over the bulk Compton and external inverse Compton processes
(see Sect. 4).

\section{Discussion}

We have shown that bulk Comptonization of the radiation from an
accretion disc around a supermassive black hole, although important in
a bolometric sense, would be currently unobservable, since its
spectrum peaks in the far UV. On the contrary, the bulk Compton
process on photons produced in the broad--line region can give raise
to an observable feature which can be represented by a blackbody.  The
entire (bulk Compton+relativistic Compton) spectral shape, can then be
approximated as blackbody+power law.

The bulk Compton feature is however difficult to detect, even in the
most favorable case. Firstly, in general it can be a transient
feature, if the plasma injection or dissipation in the jet is not
continuos. Such possibilities are in general supported by the strong
variability across the electromagnetic spectrum of blazars. Above we
mimicked these possibilities assuming that the dissipation region is a
shell moving along the jet, thus giving raise to a bulk Compton
feature when the shell is close to the BLR. However, it is well
possible that a 'cold' component is in fact more continuous. In such a
case the bulk Compton component would correspond to the `integration'
of the spectra shown. Secondly, the radiation has to be strongly
beamed, namely the jet has to be highly relativistic and the angle
with the line of sight small.  A third condition for a bulk Compton
feature to be detectable is that other emission processes do not mask
it. The most likely component that could hide it is that produced by
SSC. It should be noted that the estimated spectrum (see Fig. 4) does
not include emission via the SSC process, as this would require to
assume -- as another free parameter -- the relative intensity of the
external soft photon and magnetic fields.  The SSC could in principle
provide a significant contribution at the UV and softest X-ray
energies even in powerful sources. However, from the observational
point of view, the SSC component contributes the least to the X--ray
spectrum of the most powerful sources, since among blazars they show
the flattest X--ray spectra and the largest ``Compton dominance",
i.e. the ratio of the high vs low energy component increases as the
bolometric luminosity does (Fossati et al. 1998).  This can been
interpreted (Ghisellini et al. 1998) as due to the increased
importance of the external seed radiation field relative to the
locally produced synchrotron one with increasing source power.

This contrasts with the finding that the radiation energy density of
BLR photons decreases or remains constant in higher luminosity
objects, according to the relations reported by Kaspi et al. (2000)
and Bentz et al. (2006), respectively.  On the other hand, BLR photons
might not be the only contributors to the seed field: the BLR clouds
themselves and/or some intra-cloud scattering material can enhance and
isotropize beamed synchrotron radiation from the jet, the external
radiation (Ghisellini \& Madau 1996) and disc radiation (Sikora,
Begelman \& Rees 1994).  The presence and the optical depth of such
scattering material could depend on the disc luminosity: more massive
outflows or winds from the disc could be favoured at accretion rates
near the Eddington one.  The emission lines and thermal optical--UV
component in high redshift blazars indeed indicate accretion
luminosities close to the Eddington ones even for black hole masses
exceeding $10^9 M_\odot$.

If the Lyman--$\alpha$ hydrogen line provides most of the seed photon
field, the detection of the bulk Compton signature would constitute a
powerful diagnostic to estimate the product $\delta \Gamma$.  Another
independent estimate of $\Gamma$ or $\delta$, would then allow to
estimate both the jet speed and viewing angle. For instance, if the
source shows superluminal motion with an apparent speed $\beta_{\rm
  a}$ (and assuming $\Gamma$ does not change) we have:
\begin{equation}
\Gamma \delta \, =\, {\nu_{\rm p} \over \nu_{\rm L\alpha}}
\end{equation}
\begin{equation}
\beta_{\rm a} \, =\, {\beta\sin\theta_{\rm v} 
\over 1-\beta\cos\theta_{\rm v}}
\, \to\, \Gamma\delta\, =\, {\beta_{\rm a} 
\over \beta\sin\theta_{\rm v}}
\end{equation}
where $\nu_{\rm p}$ is the peak of the bulk Compton spectrum.  The
above two equations can be solved for $\theta_{\rm v}$ and $\Gamma$:
\begin{equation}
\sin\theta_{\rm v}\, =\, {\beta_{\rm a} \over \beta}\, \, 
{\nu_{\rm L\alpha} \over \nu_{\rm p} },
\end{equation}
\begin{equation}
\Gamma \, =\, { \nu_{\rm p}/\nu_{\rm L\alpha} \over \left[ 2 \nu_{\rm
p}/\nu_{\rm L\alpha} -1-\beta_{\rm a}^2\right]^{1/2} }. 
\end{equation}

The other key diagnostic which can be inferred from a bulk Compton
feature is the kinetic energy associated to the cold scattering
material, which is independent of the magnetic field intensity and the
filling factor of the scattering plasma.

In the case of GB B1428+4217, the above interpretation of the SED
implies $E_{\rm kin} = N \Gamma m_{\rm p} c^2 \sim 10 (3.4) 1.9\times
10^{53}$ erg, for a proton-electron plasma.  If the width of the shell
containing this mass is of the order of the initial jet size (in the
observer frame, i.e. $\Delta R\sim 6r_{g} \sim 3.6\times 10^{15}$ cm),
then the corresponding kinetic luminosity is $L_{\rm kin} = E_{\rm
kin} c/(6r_{\rm g}) \sim 8.6 (2.8) \times 10^{47}$ erg s$^{-1}$,
corresponding to about the luminosity radiated by the accretion disc.
$L_{\rm kin}$ is clearly less if there are electron--positron pairs;
it should be noticed that in the case of a pair--dominated plasma, the
jet could be significantly decelerated because of the drag due to the
bulk Compton, but even more so when dissipation occurs and also the
`rocket' effect would come into play.

Note also that $L_{\rm kin}\propto 1/\Delta R$. $\Delta R$ is here a
free, unknown parameter and indeed a different assumption on $\Delta
R$ has been made by Ghisellini et al. (2003). However, despite this
uncertainty on $L_{\rm kin}$, the effective $E_{\bf kin}$ is in fact
substantially independent of such choice.  Fitting the overall SED
would also allow us to estimate the magnetic field intensity and hence
the power in Poynting flux. To this aim we will examine the X--ray
data and SED of powerful, high redshift blazars with indications of a
bulk Compton feature (Fabian et al., in prep.). Interestingly, in
agreement with the prediction of this scenario, indications of soft
X--ray flattening appears to be connected with the source X--ray
luminosity (Yuan et al. 2006).

Future large area X--ray instruments might even provide time dependent
X--ray spectra from which in principle should be possible to constrain
the jet acceleration. Nevertheless, different shells might contribute
to the observed spectrum and thus the deconvolution from X-ray
variability to plasma acceleration might be much more complex than
outlined here.

\section{Conclusions}
The study of the bulk Compton spectrum produced by cold plasma
accelerating in a blazar jet showed that:

\noindent
1) If dissipation (i.e. acceleration/injection of relativistic
leptons) occurs during the (early) shell acceleration phase, the
radiation from the disc can provide a relevant source of photons for
the relativistic external Compton process.  Broad line radiation
becomes competitive and dominant for such process at a shell--disc
distance comparable to the broad line region.  The dominance between
the two components depends on the spectral index of the relativistic
Compton spectrum.

\noindent
2) The bulk Comptonization of the radiation field originating in the
accretion disc produces a component peaking in the far UV, thus most
difficult to observe, and its contribution to the soft X--ray band is
negligible.

\noindent
3) Instead the bulk Compton scattering of broad line photons may give
an observable feature in the soft X--ray band of the most powerful
quasars, if the dissipation takes place close to, but within, the
broad line region radius when the shell has reached high bulk Lorentz
factors (and for small viewing angles).

\noindent
4) The transient nature of the bulk Compton feature and the
requirement that other radiation processes (most notably, the
synchrotron self--Compton process) do not dominate over it, imply that
the bulk Compton component is rarely detectable.  Its absence cannot
thus be considered as evidence that jets are not matter dominated.

\noindent
5) On the other hand, the detection of this feature allows two key
quantities to be estimated: i) the bulk Lorentz factor, and ii) the
amount of cold leptons in the shell and in turn the corresponding
kinetic power.

\noindent
6) In principle the time dependent behavior of the bulk Compton
component would trace the jet acceleration profile.

\noindent
7) The bulk Compton contribution would result in a flattening of the
X--ray spectrum towards softer energies, which could mimic
absorption, and possibly a hump. Indications of such features have
been found in the spectra of some powerful high redshift blazars.  An
re-analysis of their X--ray spectra and SED will be presented in a
forthcoming paper.

\section*{Acknowledgements}
We thank the referee, Rafal Moderski, for his careful reading of the
manuscript and his constructive comments, and F. Tavecchio for useful
discussions. AC and ACF acknowledge the Italian MIUR and the Royal
Society for financial support, respectively.

\end{document}